\documentclass[utf8]{article}

\usepackage{geometry}
\usepackage{
    url,
    hyperref,
    lineno,
    amsfonts,
    amssymb,
    amsmath,
    bm,
    tikz
}
\usepackage[onehalfspacing]{setspace}
\usepackage[euler]{textgreek}

\usetikzlibrary{shapes, arrows, positioning}
\tikzset{
    block/.style={
        draw,
        rectangle,
        minimum height=3em,
        minimum width=6em
    },
    sum/.style={
        draw,
        circle
    }
}

\def\mxb#1{\bm{\mathrm{#1}}}

\title{
    Delayed closed-loop neurostimulation for the treatment of pathological brain rhythms in mental disorders
}

\author{ Thomas Wahl\footnote{
ICube, MLMS, University of Strasbourg; MIMESIS Team, Inria Nancy - Grand Est, Strasbourg, France}, Joséphine Riedinger$^{*,}$\footnote{INSERM U1114, Neuropsychologie cognitive et physiopathologie de la schizophrénie, Strasbourg, France}
, Michel Duprez$^{*}$, Axel Hutt$^{*}$}

\begin{document}

\maketitle

\begin{abstract}

Mental disorders (MD) are among the top most demanding challenges in world-wide health. According to the World Health Organization, the burden of MDs continues to grow with significant impact on health and major social and human rights. A large number of MDs exhibit pathological rhythms, which serve as the disorders characteristic biomarkers. These rhythms are the targets for neurostimulation techniques. Open-loop neurostimulation employs stimulation protocols, which are rather independent of the patients health and brain state in the moment of treatment. Most alternative closed-loop stimulation protocols consider real-time brain activity observations but  appear as adaptive open-loop protocols, where e.g. pre-defined stimulation sets in if observations fulfil pre-defined criteria. The present theoretical work proposes a fully-adaptive closed-loop neurostimulation setup, that tunes the brain activities power spectral density (PSD) according to a user-defined PSD. The utilized brain model is non-parametric and estimated from the observations via magnitude fitting in a pre-stimulus setup phase. Moreover, the algorithm takes into account possible conduction delays in the feedback connection between observation and stimulation electrode. All involved features are illustrated on pathological \textalpha- and \textgamma-rhythms known from psychosis. To this end, we simulate numerically a linear neural population brain model and a non-linear cortico-thalamic feedback loop model recently derived to explain brain activity in psychosis.


\end{abstract}

\textbf{Keywords:} neurostimulation, closed-loop, control, real-time, delay, EEG

\section{Introduction}

Electrical neurostimulation is an old human idea, and has been a well-established therapy for mental disorders for few decades. Caius Plinius during Antiquity and Scribonius Largus, who lived in the first century AD, proposed respectively contacts with the Electric ray (Torpedo Fish) for the treatment of post-partum pain and severe headaches. In the 19th century, electrical stimulation was commonly prescribed by neurologists for nervous disease \cite{Edel_Caroli}. Today, various electrical stimulation techniques exist to modulate neuronal systems and novel techniques for an optimal clinical treatment of a specific pathology gain more and more attention. They could be used as an additional therapeutic lever or as an alternative to pharmacological medication, thus representing a hope for pharmaco-resistant forms of disease.
 
Brain oscillations result from coordinated electrical neuronal tissues activity within and between structures and networks. Implicated in various neural processes, such as perception, attention and cognition, their disruption yields pathological rhythms, which reflect abnormal activity of the implicated brain network, notably at the cellular and molecular level~\cite{Basar}. These pathological rhythms serve as good biomarkers for neuropathologies. For instance, neurophysiological studies have revealed that a large number of mental disorders exhibit  pathological rhythms, which do not occur in healthy patients~\cite{Schulman}.
Neurostimulation techniques have identified such pathological rhythms as good stimulation targets for the treatment of brain oscillatory disorders. 
Neurostimulation induces electric currents in neuronal tissue. Depending on the stimulation protocol, i.e. the temporal stimulation current shape, its duration and pause and the number of repetitions, neurostimulation can lead to neural plasticity effects or to pacemaker-like brain stimulation, respectively.

For example, Deep Brain Stimulation (DBS) is an invasive technique and proposed for patients suffering from severe pharmaco-resistant Parkinson's disease (PD) or obsessive-compulsive disorders. In PD patients aberrant hypersynchronicity and hyperactivity in the \textbeta-frequency band (12-30 Hz) of the basal ganglia-thalamocortical network can be addressed by the pharmacological medication (e.g. Levodopa) or DBS. The conventional DBS protocols focus on the subthalamic nucleus or globus pallidus stimulation 

continuously at a temporally constant frequency about 130 Hz. The suppression of the pathological beta oscillations was correlated with improving motor symptoms \cite{Kuhn_2008}. Recent  techniques~\cite{Fleming,Hosain} propose to apply an adaptive closed-loop stimulation protocol based on observed intracranial brain activity. In addition to this intracranial neurostimulation technique, transcranial electrical stimulation (TES) and transcranial magnetic stimulation (TMS) are  non-invasive neuromodulation approaches in which, respectively, a low electrical current and a magnetic field are applied to the cortical tissues. The TES current modalities include direct currents (tDCS), i.e. constant currents, alternating current (tACS), i.e. typically oscillatory currents, and random noise-shape currents (tRNS), which typically includes frequencies above the \textbeta-frequency band.
It was shown that tDCS can improve cognitive performance in healthy subjects~\cite{Brunelin} and patients~\cite{Antal} and it is applied as a therapeutic means to target brain network dysfunctions, such as Attention-Deficit/Hyperactivity Disorder~\cite{Nejati} and major depressive disorder~\cite{Bennabi}.

Although the neurostimulation techniques mentioned above may permit to alleviate mental disorder patients from symptoms, the success rate of these treatments is still limited~\cite{NASR2022102311}. This underperformance results from non-optimal choices of the stimulation protocol originating from the lack of understanding of the underlying neural response to stimulations and the non-patient specific stimulation protocol. In other words, typically the stimulation protocol (including size, duration, repetition cycle of the stimulation signal) is open-loop, i.e. pre-defined without taking into account the current brain/health state of the patient~\cite{tES2011}. This non-optimal approach is inferior to so-called closed-loop techniques, which adapt to the patients current brain/health state. Such an adaptive, or closed-loop, approach has been introduced for intracranial~\cite{Prosky,Stanslaski,Hartshorn} and transcranial stimulation~\cite{TERVO2022523}. Recently proposed closed-loop methods are adaptive in the sense that a pre-defined stimulation signal is applied when observed brain activity fulfills certain criteria, such as passing an amplitude or power threshold. While this adaptive approach improves existing open-loop methods, the pre-defined stimulation signal may still be non-optimally chosen.

We propose to estimate a stimulation signal on the basis of observed brain activity. The target stimulation signal is not pre-defined as in the open-loop setting but computed according to a pre-defined target spectral power distribution of the brain activity. To our best knowledge, this focus on a target brain activity spectral distribution has not been proposed before in a closed-loop neurostimulation setup. We argue that it is the natural choice for a closed-loop optimization in the presence of pathological rhythms: typically the pathology is identified by an abnormal power in a certain frequency band and the closed-loop control aims to modify this power value in such a way that the final brain activity power spectral  distribution resembles the distribution of a healthy subject. This approach implies the hypothesis that modifying the observed pathological brain rhythms of a patient to resemble brain rhythms of a healthy subject renders the patients brain state and improves the patients health situation. This assumption was motivated by the impressive improving impact of DBS in psychiatric disorders~\cite{Holtzheimer}.

Technically, the proposed method aims to reshape the spectral distribution of observed data, such as electroencephalographic data (EEG). For illustration, we consider pathological brain rhythms observed in  psychosis in the \textalpha-~\cite{howells2018electroencephalographic} and \textgamma-band~\cite{schbul2015}. 
 Our method relies on the extraction and the filtering in real-time of the brain resting state activity signal, using the EEG and an estimated brain response model. The underlying brain model is fully non-parametric and estimated from observed resting state EEG. Moreover, we consider the fact that the closed-loop feedback exhibits a certain conduction delay between measurement and stimulation. This conduction delay results from the transmission delay in the hardware and the numerical computation time of the stimulation signal. Very first estimates of this delay time are in the range of few tens of milliseconds~[Private communication, Isope, 2020], i.e. in the range of EEG signal time scales. Consequently, the present feedback delay in real-world systems may affect the methods performance. To our best knowledge, the present study is the first considering delays in closed-loop neurostimulation systems.

The remaining article is organized as follows :
Section \ref{sec:material_and_methods} presents the neurostimulation setup and the closed-loop circuit studied in the rest of this paper. Then, we propose a model-based controller design to apply desired modifications to the observed activity signal. Subsequently, we propose a model estimation method to extract the brain input response model needed for the controller design. Later, we address the problem of the closed-loop delay by designing an additional system to approximate the future values of the observations. Finally, we present two brain models, which illustrate and validate the proposed method.
Then, Section \ref{sec:results} presents the simulation results of our circuits, including the accuracy of the model estimation step and the delay compensation.
Lastly, in section \ref{sec:discussion}, we discuss the results of the method presented in the paper compared to the state of the art, mention limitations and pinpoint some perspectives and possible experimental tests.

\section{Material and methods}
\label{sec:material_and_methods}

\subsection{Neurostimulation setup}

We build a theoretical plant as a circuit containing a stimulation element and an observation element, both connected to the model brain system under study. In real practice, the stimulation element corresponds to the neurostimulation device, such as a TES system or a TMS coil. In contrast, the observation element may represent electro-/magneto-encephalographic electrodes (in the following called EEG) or electrodes observing Local Field Potential. We define the time-dependent functions $u:\mathbb{R}\rightarrow\mathbb{R}$ and $y:\mathbb{R}\rightarrow\mathbb{R}$ as the input stimulation current and the output EEG signal, respectively.

If no input current is applied,  the output is a non-zero stochastic signal $y_0$ corresponding to the measured  resting state EEG activity  and a non-zero neurostimulation current alters the output signal as a linear response. This alteration is caused by a change in the brain activity in response to the neurostimulation input and a direct measurement of the input current. The latter is undesirable as it is not correlated with brain dynamics but only with neurostimulation and measurement devices. In the following, we assume that observations include brain dynamics correlated output only while direct current measurements are filtered out. A method to remove the direct current measurement from the EEG signal is discussed in Section \ref{sec:remove_interaction}.

Then, we define the  plant $\mathcal{P}$ as the system that takes $u$ as its input and generates an output $y$ which is equal to $y_0$ when no input is applied. By modeling the dynamics of $\mathcal{P}$, our goal is a neurostimulation signal $u$ that causes predetermined changes in the spectral power amplitude of the output signal  $y$. In our case, the goal is to increase the activity in the alpha band ($8-12$Hz) and decrease the activity in the gamma band ($25-55$Hz).

\subsection{Linear time invariant model}

We assume that the observed output response to a small neurostimulation input $u$ is linear and time-invariant (LTI). This assumption is supported by multiple results across literature \cite{Liu2010, Popivanov1996, Kim2016}. Thus, there is an underlying LTI system $\mathcal{G}$ that produces an output $y_u$ for any given input $u$. For this system, we can define a function $g:\mathbb{R}\rightarrow\mathbb{R}$, which is the output produced by the plant input response system $\mathcal{G}$ in response to a unit impulse signal $\delta(t)$. This function $g$ is also called the unit impulse response of $\mathcal{G}$ and we have

\begin{equation*}
    y_u(t) = g(t)*u(t) := \int_{-\infty}^{+\infty}g(t')u(t - t')dt'.
\end{equation*}
with time $t$ and  $*$ denotes the convolution over time.
It leads to the total plant output

\begin{equation}
        y(t) = y_0(t) + y_u(t) 
             = y_0(t) + g(t)*u(t).
    \label{eq:output}
\end{equation}

With this choice of model, the contribution of the neurostimulation response to the total output is purely additive, allowing us to focus the analysis on $\mathcal{G}$, which represents the neurostimulation response part of the plant system. We also see that $y_0$, the resting state activity, contains the stochastic part of the output, while $y_u$ can be predicted for any known input signal $u$ if we have a model for the system $\mathcal{G}$. A method to estimate the plant input response model $\mathcal{G}$ is presented in section {\ref{sec:model_estimation}}.

\subsection{Closed-loop control}

In this section, we suppose that the function $g$ is known. The estimation of $g$ will be the aim of section \ref{sec:model_estimation}.

\begin{figure}
    \centering
    \begin{tikzpicture}[auto,>=latex']
        \node [block] at (2, 0) (ctr) {$K$};
        \node [block] at (2, 2) (sys) {$G$};
        \node [sum]   at (4, 2) (sum) {};
        \draw (sum) node [above left] {$+$};
        \draw [->] (4, 3) -- node {$y_0$} (sum);
        \draw [->] (sys) -- (sum);
        \draw [->] (5, 2) |- (ctr);
        \draw [->] (ctr) -- (0, 0) |- node {$u$} (sys);
        \draw [->] (sum) -- node {$y$} (6, 2);
        \draw [dashed] (.5, 1) rectangle (4.75, 3.25);
        \draw (2.5, 3) node {Plant};
    \end{tikzpicture}
    \caption{
    {\bf Closed-loop neurostimulation circuit}
    }
    \label{diag:circuit}
\end{figure}
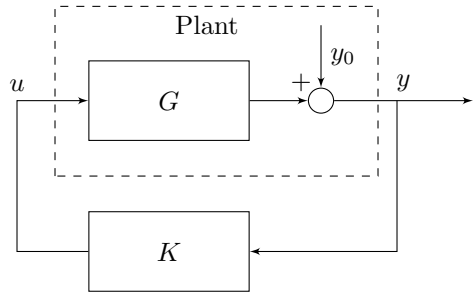

To close the loop, we generate the plant input signal $u$ as the output of a linear controller $\mathcal{K}$ in response to the plant output $y$
\begin{equation*}
    u(t) = k(t)*y(t),
\end{equation*}
where $k:\mathbb{R}\rightarrow\mathbb{R}$ is the unit impulse response of the controller $\mathcal{K}$. We can now rewrite Eq.~(\ref{eq:output}) as
\begin{equation}
    y(t) = y_0(t) + g(t)*k(t)*y(t).
    \label{eq:closed-loop}
\end{equation}
Here, we assume that no  delay between observation and stimulation application is present. We will relax this condition in section~\ref{subsec_delay}. To solve Eq.~(\ref{eq:closed-loop}), we apply the Laplace transform defined for each time-dependent function $x:\mathbb{R}\rightarrow\mathbb{R}$ by
\begin{equation}
    X(s) = \mathcal{L}\{x(t)\}(s) := \int_{0^-}^{+\infty}x(t)e^{-st}dt,
    \label{eq:laplace}
\end{equation}
Thus, we define $Y:\mathbb{C}\rightarrow\mathbb{C}$, $Y_0:\mathbb{C}\rightarrow\mathbb{C}$, $G:\mathbb{C}\rightarrow\mathbb{C}$ and $K:\mathbb{C}\rightarrow\mathbb{C}$ as the Laplace transforms of respectively $y$, $y_0$, $g$ and $k$, allowing us to write Eq.~(\ref{eq:closed-loop}) as
\begin{equation*}
    Y(s) = Y_0(s) + G(s)K(s)Y(s).
\end{equation*}
Hence
\begin{equation}
    Y(s) = \frac{1}{1 - G(s)K(s)}Y_0(s).
    \label{eq:closed-loop-output}
\end{equation}
We now have an equation for the closed-loop output in function of the resting state activity. A block diagram of the closed-loop circuit is shown in Fig.~\ref{diag:circuit}.
Hence to design the frequency distribution of $y$ we tune the frequency distribution of the transfer function $K$ of the controller $\mathcal{K}$

\subsubsection*{Controller synthesis}

Our closed-loop setup aims to tune the observation power spectrum, or equivalently, the choice of $Y(s)$ subjected to the resting state $Y_0(s)$. To this end, 
we define a linear filter $\mathcal{H}$ with transfer function $H:\mathbb{C}\rightarrow\mathbb{C}$  and

\begin{equation}
    Y(s) = Y_0(s) + H(s)Y_0(s).
    \label{eq:target_output}
\end{equation}
Specifically, we intend to restore the physiological state of the brain, e.g. of a schizophrenic patient as our motivation, with an observed EEG presenting low alpha activity and high gamma activity. The chosen filter $\mathcal{H}$ is a weighted double bandpass filter with positive weight in the \textalpha-frequency band to increase \textalpha-power and negative weights in the \textgamma-band to decrease the systems \textgamma-activity. The filter's transfer function is defined as

\begin{equation*}
    H(s) = c_1 \frac{2\pi B_1 s}{s^2 + 2\pi B_1 s + (2\pi f_1)^2} + c_2 \frac{2\pi B_2 s}{s^2 + 2\pi B_2 s + (2\pi f_2)^2}.
\end{equation*}

The exact parameters of $\mathcal{H}$ are shown in table \ref{table:filter}.

We can synthesize the closed-loop controller $\mathcal{K}$, by combining equations (\ref{eq:closed-loop-output}) and (\ref{eq:target_output}) and solving for $K$ as

\begin{eqnarray}
\frac{1}{1 - G(s)K(s)}Y_0(s) &=& Y_0(s) + H(s)Y_0(s)\nonumber\\
          K(s) &=& \frac{H(s)}{(1 + H(s))G(s)}.\label{eqn:K}
\end{eqnarray}

Therefore, if we know the plant input response transfer function $G$, we can find that desired controller transfer function $K$ by Eq.~(\ref{eqn:K}). Once the transfer function is obtained, we can use it to find a corresponding state-space representation \cite{hespanha2018linear} for time domain simulations.

\begin{table}
    \centering
    \begin{tabular}{lll}
        \hline\noalign{\smallskip}
        parameter & description & value  \\
        \noalign{\smallskip}\hline\noalign{\smallskip}
        $f_1$ & \textalpha-band natural frequency  & 10ms \\
        $B_1$ & \textalpha-band width              & 4Hz \\
        $c_1$ & \textalpha-band weight             & 1.0\\
        $f_2$ & \textgamma-band natural frequency & 40ms \\
        $B_2$ & \textgamma-band width             & 30Hz \\
        $c_2$ & \textgamma-band weight            & -0.5\\
        \noalign{\smallskip}\hline
    \end{tabular}
    \caption{
        {\bf Parameter set of the filter} $\mathcal{H}$. The frequency parameters are chosen based on the alpha frequency range (8-12Hz) and the gamma frequency range (25-55Hz) in an EEG. The weighting parameters $c_1$ and $c_2$, respectively positive and negative, corresponding to the choice to increase the alpha activity and decrease the gamma activity.
    }
    \label{table:filter}
\end{table}

\subsection{Model estimation}
\label{sec:model_estimation}
The design of our closed-loop controller requires estimating the plant input response system $\mathcal{G}$, which in practice includes the brain dynamics, the neurostimulation device and the observation device. Our approach  includes the estimation of $\mathcal{G}$  directly from observed brain activity, such as EEG of the patient. This ensure that the estimated plant model will be as close as possible to the real brain dynamics in the corresponding experimental conditions. To this end,  we first need to find a way to measure the plant input response without also measuring the plant  resting state activity. This is not trivial since the observed signal is the sum of the resting state activity and the stimulation response.

\subsubsection*{Signal extraction}

Let us consider an open-loop setup with an arbitrary input $u$ applied to the plant, which generates  the output described by Eq.~(\ref{eq:output}). In this equation, we only know $u$ and $y$, and want to estimate the  impulse response $g$. The problem is that we cannot observe $y_0$ only during the stimulation. Hence, based on previous data recordings, we need to find a way to predict the dynamics of $y_0$ during the stimulation.

First, we provide the following standard definitions that are important in the subsequent discussion.  For any time domain signal $x:\mathbb{R}\rightarrow\mathbb{R}$, we denote the Fourier transform by

\begin{equation}
    \hat{x}(f) = \mathcal{F}\{x(t)\}(f) := \int_{-\infty}^{\infty}x(t)e^{-2\pi ift}dt.
    \label{eq:fourier}
\end{equation}

We define $\alpha_0:\mathbb{R}\rightarrow\mathbb{R}$ and $\alpha_u:\mathbb{R}\rightarrow\mathbb{R}$ such as $\alpha_0(t) = y_0(t) - \bar{y}_0$ and $\alpha_u(t) = y_u(t) - \bar{y}_u$ where $\bar{y}$, $\bar{y}_0$ and $\bar{y}_u$ are respectively the ensemble means of $y$, $y_0$ and $y_u$. \\

We assume  that $y_0$ is a wide-sense-stationary (WSS) random process, i.e. its mean and variance do not depend on time. According to the Wiener-Khinchin theorem~\cite{Khinchin,Gardiner04}, the autocorrelation function of a wide-sense-stationary random process has a spectral decomposition given by the power spectrum of that process

\begin{equation*}
S_{yy}(f) = |\hat{\alpha}(f)|^2,
\end{equation*}

where $\hat{\alpha}:\mathbb{R}\rightarrow\mathbb{C}$ is the Fourier transform of $\alpha(t) = y(t) - \bar{y}\in \mathbb{R}$ and  $S_{yy}:\mathbb{R}\rightarrow\mathbb{R^+}$ is the spectral density of $y$.

Then, we can write Eq.~(\ref{eq:output}) as
\begin{equation*}
    \bar{y} + \alpha(t) = \bar{y}_0 + \alpha_0(t) + \bar{y}_u + \alpha_u(t),
\end{equation*}
where $\bar{y} = \bar{y}_0 + \bar{y}_u$. 
The equation then simplifies to
\begin{equation*}
    \alpha(t) = \alpha_0(t) + \alpha_u(t).
\end{equation*}
By application of the Fourier transform, we obtain
\begin{equation*}
    \hat{\alpha}(f) = \hat{\alpha}_0(f) + \hat{\alpha}_u(f)
\end{equation*}
and
\begin{equation*}
    |\hat{\alpha}(f)|^2 = |\hat{\alpha}_0(f)|^2 + |\hat{\alpha}_u(f)|^2 + 2\mathrm{Re}[\hat{\alpha}_0(f)\hat{\alpha}_u(f)^*].
\end{equation*}

In the following, we compute the ensemble average of each term of this equation. Since $\alpha$ and $\alpha_u$ are two independent processes sampled at different times and $\langle \hat{\alpha}_0 \rangle=\langle \hat{\alpha}_u\rangle=0$.

Hence
\begin{eqnarray}
    \langle2\mathrm{Re}(\hat{\alpha}_0(f)\hat{\alpha}_u(f)^*)\rangle = 2\mathrm{Re}[\langle\hat{\alpha}_0(f)\hat{\alpha}_u(f)^*\rangle]\nonumber
    =0.\label{eqn:nocorr}
\end{eqnarray}
Here and in the following, $\langle\cdot\rangle$ denotes the ensemble average. We point out that although Eq.~(\ref{eqn:nocorr}) does hold when considering the ensemble average of the signals, fluctuations around $0$ still remain in Eq.~(\ref{eqn:nocorr}) for finite ensemble number of finite time signals.

Nevertheless, this yields
\begin{equation}
    \begin{split}
    \langle|\hat{\alpha}_u(f)|^2\rangle &= \langle|\hat{\alpha}(f)|^2\rangle - \langle|\hat{\alpha}_0(f)|^2\rangle. \\
    \end{split}\label{alpha_u}
\end{equation}

Using Eq.~(\ref{eq:output}), we can express $\hat{\alpha}_u$ in terms of the input impulse response $g$ and the input $u$

\begin{equation}
\label{eqn_alphau2}
    \begin{split}
        \hat{\alpha}_u(f) &= \mathcal{F}\{y_u(t) - \bar{y}_u\}(f) \\
                          &= \mathcal{F}\{g(t) * [u(t) - \bar{u}]\}(f) \\
                          &= \hat{g}(f)\mathcal{F}\{u(t) - \bar{u}\}(f)~.
    \end{split}
\end{equation}
This equation permits to estimate the transfer function $\hat{g}$, see Section~\ref{sec:results}.

To express the transfer function $\hat{g}$ in Laplace space, we use the fact that a unit impulse response function is non-zero only for positive time values $t$. Hence, based on equations (\ref{eq:laplace}) and (\ref{eq:fourier}), for $s = 2\pi if$, we can write the Laplace transform $G$ as
\begin{equation*}
    G(2\pi i f) = \int_{0^-}^{+\infty}g(t)e^{-2\pi ift}dt = \int_{-\infty}^{+\infty}g(t)e^{-2\pi ift}dt = \hat{g}(f).
\end{equation*}

We now need a method to generate a LTI system with a transfer function that matches the magnitude data computed with the formula. This is achieved by the magnitude vector fitting algorithm.

\subsubsection*{Magnitude vector fitting}

Our goal is now to find a transfer function $G$ corresponding the magnitude data $|\hat{g}(f)|^2$. For this purpose, we use a variant of the vector fitting algorithm design to work even with only the magnitude data. This method is called magnitude vector fitting \cite{de2010robust}.

It allows to fit a passive LTI system to data by fitting the model transfer function. The system is synthesized such that the mean square error between the magnitude data sample and the transfer function evaluated at the same frequency points is minimized. \cite{de2010robust} show that the transfer function of the fitted model reproduces both the magnitude and the phase shift of the original transfer function, although the fitting has been performed using sampled magnitude data only.

By minimizing the mean square error, the algorithm ensures that the transfer function of the fitted model accurately matches the original model as represented by the reconstructed gain data. Furthermore, to assess the accuracy of the reconstruction, we also compare the fitted model to the transfer function of the linearized brain model used for the simulation. This allows to double-check the validity of the reconstructed magnitude and also to verify if the reconstructed phase fits the phase of the original model as closely as possible cf. Fig~\ref{fig:psdfit}C,D.

\subsection{Delay compensation}\label{subsec_delay}

Realistic feedback loops exhibit conduction delays between the moment of observation and feedback stimulation. Reasons for such delays are finite conduction speeds in cables, electronic switches, interfaces and delays caused by the controller device to compute numerically adapted stimuli. In systems with large time scales, such as controlled mechanical devices on the centimeter or larger scale, such delays may be negligible. Conversely biological systems such as the brain evolve on a millisecond scale and conduction delays may play an important role. Preliminary estimation of input and output devices of desktop computers have revealed an approximate delay of $\sim 10$ms. By virtue of such delays, it is important to take them into account in the closed-loop between the moment of observation and stimulation.

The different sources of delay can be represented as plant input and output delays. Since the controller $\mathcal{K}$ is LTI, the input and output delays can be concatenated into one single plant input delay. Hence, in our setup, we model the delay as an input delay $\tau$ in the system $\mathcal{G}$, modifying $y(t) = g(t) * u(t)$ in Eq.~(\ref{eq:output}) to $y(t) = g(t) * u(t - \tau)$. The Smith predictor~\cite{smith1959} \cite{morari1989} is a known method to compensate such delay times. However, in the present problem, this approach allows controlling a limited frequency band only (see Fig.~\ref{fig:compare}A)). Consequently, it was necessary to invent another method. Since the plant input $u$ is generated by the controller $\mathcal{K}$, we modify the controller to compensate the delay. To this end, the new controller $\mathcal{K}$ is chosen to estimate  the future value of $u$ instead of the present value.  A method to apply this controller modification is presented in Section \ref{sec:results:delay_compensation}.

\subsection{Brain models}

Our closed-loop control method works for any LTI brain model. Furthermore, we want to show that it also produces good results on non-linear brain models, for which the neurostimulation input response behaves closely to an LTI system, when the input is sufficiently small. To this end, we present two models used to test our method. The first one is a linear neural population model of cortical activity, and the second one is a non-linear cortico-thalamic neural population model with cortico-thalamic delay.

\subsubsection{Linear brain model}\label{subsubsection_linear}

We describe neural population activity with a noise-driven linear model \cite{hutt2013}. The model is composed of two pairs of interacting excitatory and inhibitory populations. Here we have $V_{e,i}^{(1,2)}:\mathbb{R}\rightarrow\mathbb{R}$, representing the mean activity of the associated population, where $V_e^{(1,2)}$ and $V_i^{(1,2)}$ correspond respectively to excitatory and inhibitory populations. Each population is driven by noise $\xi_{1, 2}:\mathbb{R}\rightarrow\mathbb{R}$ and the external input $u:\mathbb{R}\rightarrow\mathbb{R}$, according to the following differential equations:

\begin{equation}\left\{
    \begin{array}{rl}
        \tau_{e,1}\frac{d{V_e}^{(1)}(t)}{dt} &= (-1 + N_{11})V_e^{(1)}(t) - N_{11}V_i^{(1)}(t) + b_1 u(t) + \xi_{1}(t), \vspace*{2mm}\\
        \tau_{i,1}\frac{d{V_i}^{(1)}(t)}{dt} &= N_{21}V_e^{(1)}(t) + (-1 - N_{21})V_i^{(1)}(t) + b_2 u(t), \vspace*{2mm}\\
        \tau_{e,2}\frac{d{V_e}^{(2)}(t)}{dt} &= (-1 + N_{12})V_e^{(2)}(t) - N_{12}V_i^{(2)}(t) + b_3 u(t) + \xi_{2}(t),  \vspace*{2mm}\\
        \tau_{i,2}\frac{d{V_i}^{(2)}(t)}{dt} &= N_{22}V_e^{(2)}(t) + (-1 - N_{22})V_i^{(2)}(t) + b_4 u(t), \\
    \end{array}\right.
    \label{eq:linear_neuropop}
\end{equation}

where the noise $\xi_{1,2}$ is uncorrelated Gaussian distributed with zero mean and variance $\kappa_{1,2}^2=10^{-7}$, and the stimulation $u$ is weighted by the coupling constants $b_i > 0$ of the corresponding population. In addition, $\tau_{(e,i),(1,2)}$ are the synaptic time constants of the populations, and constants $N_{ij} > 0$ are interaction gains of the respective population. Table~\ref{table:parameters} provides the parameters employed in subsequent simulations.

The observed output
\begin{equation*}
    y(t) = V_e^{(1)}(t) -V_i^{(1)}(t) + V_e^{(2)}(t) - V_i^{(2)}(t)
\end{equation*}
is a sum of the effective field potential $V_e^{(j)}-V_i^{(j)}$ of both populations $j=1,2$, cf. Fig.~\ref{fig:compare} (top panels).

The simulation of the linear brain model in time domain is done  using the library \texttt{control} of \texttt{python}. The numerical integration is computed thanks to matrix exponential \cite{van1978}, with a simulation sampling time of $1$ms.

\begin{table}
    \centering
    \begin{tabular}{lll}
        \hline\noalign{\smallskip}
        parameter & description & value  \\
        \noalign{\smallskip}\hline\noalign{\smallskip}
        $\tau_{e,1,2}$ & exc. synaptic time constant & 5ms \\
        $\tau_{i,1,2}$ & inhib. synaptic time constant & 20ms \\
        $N_{11}$ & first exc. linear coefficient & 1.15  \\
        $N_{21}$ & first inhib. linear coefficient & 0.63  \\
        $N_{12}$ & second exc. linear coefficient & 2.52  \\
        $N_{22}$ & second inhib. linear coefficient & 6.6  \\
        $N$ & number of neurons & 1000 \\
        $\kappa^2_{1, 2}$ & noises variances & $10^{-4}/N$ \\
        $b_{1, 2}$ & input coupling constants & 0.18 \\
        $b_{3, 4}$ & input coupling constants & 0.14 \\
        \noalign{\smallskip}\hline
    \end{tabular}
    \caption{
        {\bf Parameter set of model (\ref{eq:linear_neuropop}).}
        The choice of parameter is partially based on the paper in which it was developed 
   (see \cite{hutt2013}).  }
  
    \label{table:parameters}
\end{table}

\subsubsection{Cortico-thalamic brain model}
\label{subsebsection_cct}
A different model considers the cortico-thalamic feedback circuit \cite{riedinger2022}. It describes the cortex layers I-III and the cortico-thalamic loop between cortical layers IV-VI, the thalamic relay cell population and the reticular structure. The cortical layer I-III exhibits mean activity of excitatory cells $v$ and inhibitory cells $w$. Similarly, layer IV-VIs exhibits the mean activity $V_e$ and $V_i$ and thalamic relay cell populations the mean activity $V_{th,e}$ and $V_{th,i}$. Moreover, the reticular structure has the mean activity $V_{ret}$. The fibers between the cortex and thalamus and the cortex and reticular structure exhibit a finite conduction delay $\tau$~\cite{riedinger2022,Hashemi_etal15}. The $7$-dimensional dynamical system of the brain state ${\bf x}=(v,w,V_e,V_i,V_{th,e},V_{th,i},V_{ret})\in\mathbb{R}^7$ obeys
\begin{equation}\left\{
    \begin{array}{rl}
        \mxb{\dot{x}}(t) &= \mxb{F}(\mxb{x}(t), \mxb{x}(t - \tau)) + \mxb{\xi}(t) + \mxb{B}u(t), \\
                    y(t) &= \mxb{C}\mxb{x}(t),
    \end{array}\right.
    \label{eq:ct}
\end{equation}
where the superscript $t$ denotes transposition, ${\bf F}\in\mathbb{R}^7$ is a nonlinear vector function, ${\bf B}\in\mathbb{R}^{7\times1}$ is the input coupling matrix and $\mxb{C}\in\mathbb{R}^{1\times7}$ is the observation matrix. We mention that $\mxb{B}=(b_1,b_2,b_3,b_4,0,0,0)^t,~b_i>0$, i.e. only the cortical layers are stimulated with weights $b_i$. The observation $y$ captures the activity of the cortical excitatory populations~\cite{riedinger2022,NunezBook06} with $\mxb{C}=(c_1,0,c_3,0,0,0,0),~c_i>0$. For more details, please see the Appendix.

\begin{figure}[ht!]
\begin{center}
    \includegraphics[width=\linewidth]{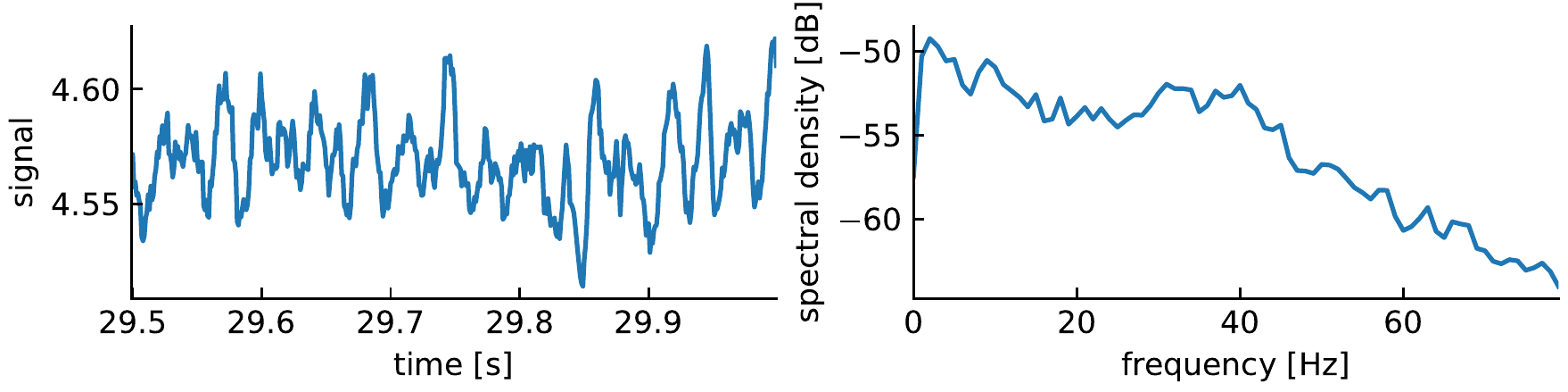}
\end{center}
\caption{
    {\bf Resting state activity computed from the cortico-thalamic brain model.} Left: Observation time series in a certain time window. Right: Power spectral density of the observation time series.
}
\label{fig:ctactivity}
\end{figure}

The time domain simulations of the cortico-thalamic model is done by numerical integration using the fourth-order Runge-Kutta method implemented by the \texttt{scipy} library in \texttt{python} with a maximum simulation time step of 1 ms. The signal produced by this cortico-thalamic brain model is shown in Fig.~\ref{fig:ctactivity}.

\section{Results}
\label{sec:results}

The present work addresses two major problems in closed-loop control: the correct model choice of the systems dynamics and the present conduction delay. The subsequent sections propose solutions for both problems and illustrate them in some detail by applying them to the linear brain activity model from section \ref{subsubsection_linear}. The final section demonstrates the closed feedback loop for the cortico-thalamic brain model from Section~\ref{subsebsection_cct}.

\subsection{Model estimation}\label{subsec_modelestimation}

Equations~\eqref{alpha_u} and \ref{eqn_alphau2} permit to 
express the magnitude of $\hat{g}(f)$ in terms of the spectral densities of observable signals

\begin{equation}
    \begin{split}
    |\hat{g}(f)|^2|\mathcal{F}\{u(t) - \bar{u}\}(f)|^2 &= |\hat{\alpha}(f)|^2 - |\hat{\alpha}_0(f)|^2 \\
    |\hat{g}(f)|^2S_{uu}(f) &= S_{yy}(f) - S_{y_0y_0}(f) \\
    |\hat{g}(f)|^2 &= \frac{S_{yy}(f) - S_{y_0y_0}(f)}{S_{uu}(f)}.
    \end{split}
    \label{eq:tf_mag}
\end{equation}

\begin{figure}[ht!]
\begin{center}
\includegraphics[width=\linewidth]{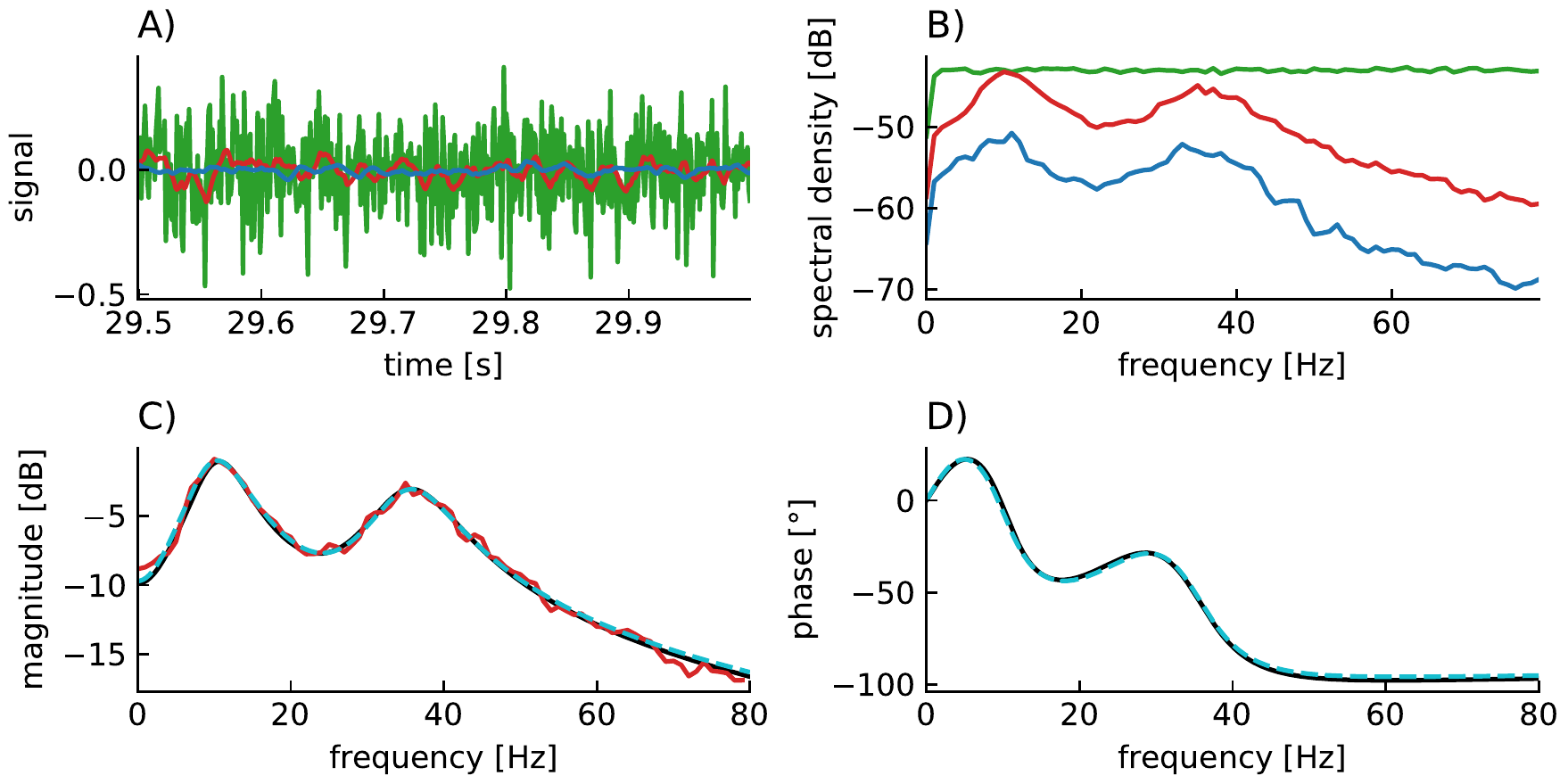}
\end{center}
\caption{
    {\bf The magnitude vector fitting algorithm successfully reconstructs the transfer function $G$ from magnitude-only data.}
    {\bf A)} Time series of the resting state activity (blue), the input signal (green) and the stimulation response (red).
    {\bf B)} Spectral densities of the simulated input signal (green), the resting state activity (blue) and the stimulation response (red). The input signal is a white noise with chosen standard deviation $0.005$.  
    {\bf C)} Reconstructed gain $|\hat{g}|$ of the plant input response. The fitted model (dashed cyan) accurately matches the original model (black). The red curve is the raw data used for fitting, computed from the spectral density data in panel A) using Eq.~(\ref{eq:tf_mag}). 
    {\bf D)} Reconstructed phase of the plant input response $\hat{g}$}.
\label{fig:psdfit}
\end{figure}

The spectral density functions $S_{y_0y_0}$ and $S_{yy}$ may be estimated numerically from output data before and during a stimulation with a known chosen stimulation function $u$. The estimation may be performed by applying conventional methods, such as the Welch method~\cite{Welch}.
These estimations provide the magnitude of the transfer function $|\hat{g}|$ by  utilizing Eq.~(\ref{eq:tf_mag}). In detail, at first, we considered the linear model (\ref{eq:linear_neuropop}) and injected a white noise current into the plant gaining the system's response signal together with the resting state activity, cf. Fig.~\ref{fig:psdfit}A. The subsequent estimation of $S_{yy}(f),~ S_{y_0y_0}(f)$ and $S_{uu}(f)$ (see Fig.~\ref{fig:psdfit}B) from the data permitted to compute the brain input response model $\hat{g}(f)$ by Eq.~(\ref{eq:tf_mag}). We observe a very good accordance of the original model response function and its estimation in magnitude (see Fig.~ \ref{fig:psdfit}C) and phase (see Fig.~ \ref{fig:psdfit}D).

\begin{figure}[ht!]
\begin{center}
    \includegraphics[width=\linewidth]{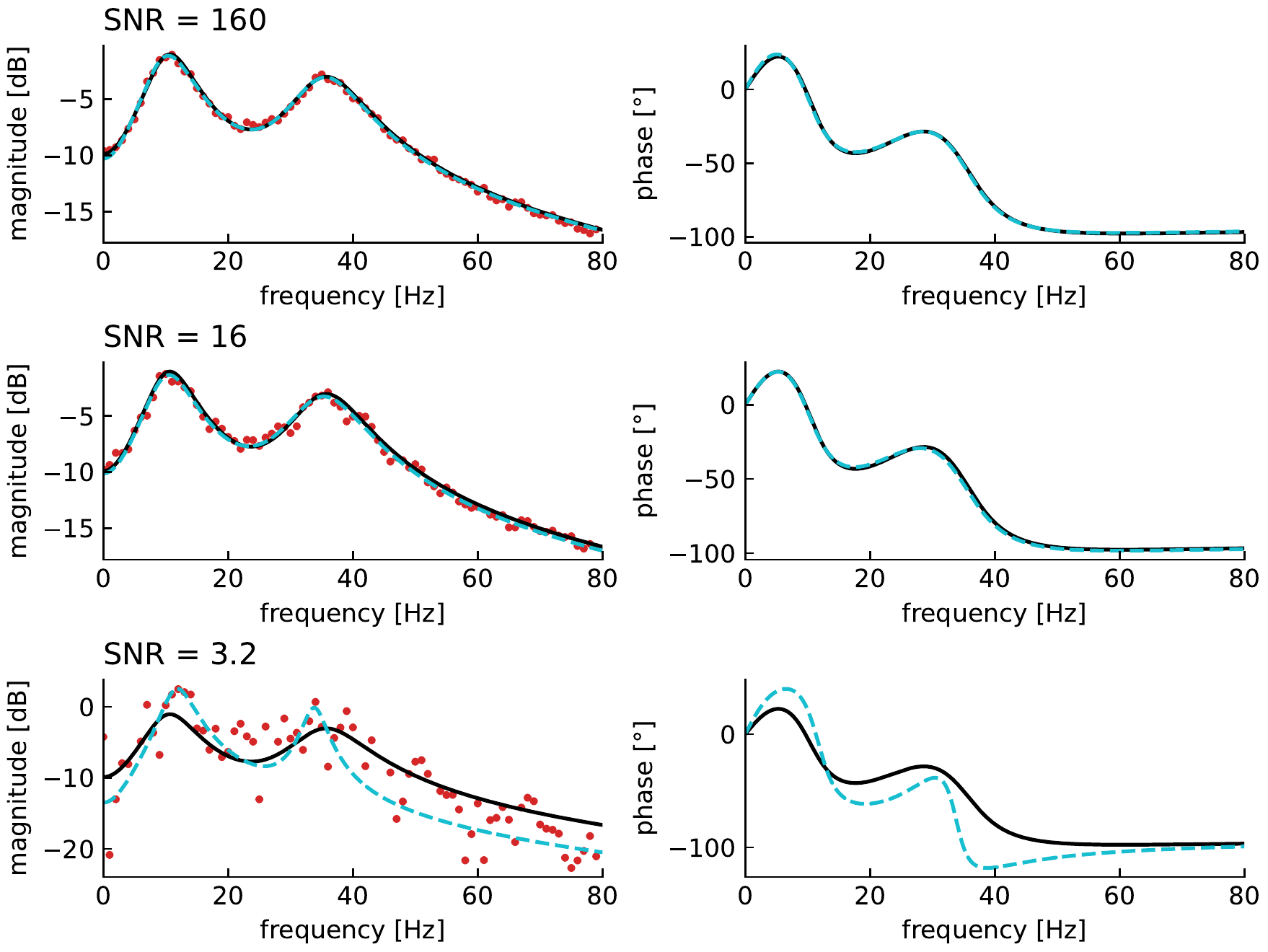}
\end{center}
\caption{
    \textbf{
        The magnitude vector fitting algorithm's performances depend on the amplitude ratio of the stimulation current and the driving noise.
    }
    Each row correspond to a different signal-to-noise ratios (SNR), computed as the ratio between the mean input coupling strength and the mean noise standard deviation. The transfer function magnitude data (red dots) are then used to synthesize a plant model via the magnitude vector fitting algorithm.
    The left (right) column corresponds to the transfer function magnitude (transfer function phase).
    We see that the noise levels in the transfer function magnitudes are higher for stronger brain-driving noise. The fitted model is coded in dashed cyan and deviates more from the original model for higher noise levels.
}
\label{fig:noise_level}
\end{figure}

The remaining error in the estimated model compared to the original model depends on the amplitude of the driving noise $\xi$, cf. Fig.~\ref{fig:noise_level}. High driving noise can also cause the magnitude vector fitting algorithm not to converge, leading to a non-minimal mean-square error between the fitted and the original models when evaluated at the frequency sample points used for the algorithm.

This problem can be solved by increasing the amplitude of the input current $u$ that we inject in the plant, which decreases the contribution of the rest state driving noise $\xi$ to the output signal relative to the input current. Although the remaining dominant input current is also noisy, its value at any time or frequency is known, meaning that it is canceled out in the ratio $\frac{S_{yy}}{S_{uu}}$ in Eq.~(\ref{eq:tf_mag}). This effectively leads to lower noise in the transfer function magnitude data extracted with Eq.~(\ref{eq:tf_mag}). The limitation is then set by the maximum amplitude of the current we are allowed to inject into the brain in a given neurostimulation setup. Indeed, the amplitude of the current is limited both for safety reasons that are beyond the scope of this paper and because of the assumption of linearity on which our method is based and which requires small currents.

\subsection{Delay compensation}
\label{sec:results:delay_compensation}

\begin{figure}
    \centering
    \begin{tikzpicture}[auto,>=latex']
        \node [block] at (0, 0) (prd) {$\Phi$};
        \node [block] at (3, 0) (ctr) {$K$};
        \node [block] at (2, 2) (sys) {$G$};
        \node [sum]   at (4, 2) (sum) {};
        \draw (sum) node [above left] {$+$};
        \draw [->] (4, 3) -- node {$y_0$} (sum);
        \draw [->] (sys) -- (sum);
        \draw [->] (5, 2) |- (ctr);
        \draw [->] (ctr) -- (prd);
        \draw [->] (prd) -- (-2, 0) |- node {$u$} (sys);
        \draw [->] (sum) -- node {$y$} (6, 2);
        \draw [dashed] (.5, 1) rectangle (4.75, 3.25);
        \draw (2.5, 3) node {Plant};
    \end{tikzpicture}
    \caption{
    {\bf Closed-loop neurostimulation circuit with predictor}
    }
    \label{diag:predictor}
\end{figure}
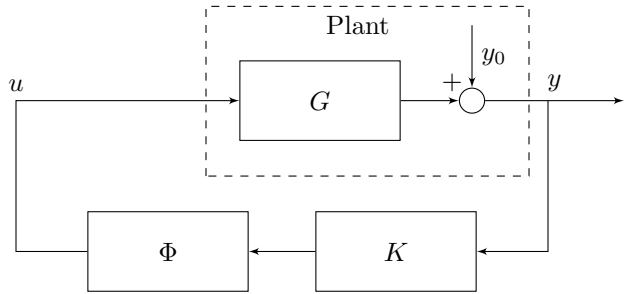

Delay compensation is achieved by adding another LTI system at the output of the controller $\mathcal{K}$ cf. Fig.~\ref{diag:predictor}, whose purpose is to reproduce the transfer function of a negative delay. We call this system the predictor $\phi$.

However, perfectly reproducing the transfer function of a negative delay would be impossible since the associated time-domain system would then be a perfect predictor, which is a non-causal, i.e. un-physical, system. Nonetheless, we can build a causal and stable system that behaves almost like a perfect predictor, however only in the frequency ranges of interest.

The numerical implementation of the controller necessitates discretization in time. Consequently, it is reasonable to choose the predictor design as a discrete-time system, meaning that for any input signal at $x_t:{\mathbb R}\to\mathbb{R}$ at an instant $t\in\mathbb{R}$, it approximately predicts the future signal $x_{t+\Delta t}$ where $\Delta t\in\mathbb{R}$ is the sampling time chosen when building the predictor. Since $x$ is a discrete sequence, its transfer function is obtained using the Z-transform, defined as

\begin{equation*}
    X(z)=\mathcal{Z}\{x_{n\Delta t}\}(z) := \sum_{n=0}^\infty x_{n\Delta t}z^{-n},
\end{equation*}
with $z\in\mathbb{C}$ and $X: \mathbb{C}\to\mathbb{C}$. Then the transfer function $\Phi:\mathbb{C}\rightarrow\mathbb{C}$ of a negative delay of one step $\Delta t$ applied to $x$ would simply be $\Phi(z) = z$, the Z-transform of a one-step delay. However, this choice would be non-causal, which is not implementable numerically in time. Nevertheless, to obtain  a stable and implementable system with a transfer function as close as possible to $z$,  we chose the ansatz
\begin{equation}
    \Phi(z_0) = \frac{b_0 z_0 + b_1}{z_0 - a} = z_0,
    \label{eq:zequation}
\end{equation}
for a fixed value $z=z_0$ and where $a\in\mathbb{R}$ is the pole of the system and $b_0\in\mathbb{R}$ and $b_1\in\mathbb{R}$ are the polynomial coefficients of the numerator of $\Phi$. This equation corresponds to the transfer function of a discrete LTI system with exactly one pole and one zero, which is the closest form of a proper rational function to the identity function of $z$ in the sense that it has only one more pole. We add the additional constraints that $|a| < 1$, since this is the necessary and sufficient condition for the discrete predictor $\mathcal{\phi}$ to be stable.

We choose to reformulate this problem by setting $a$ as a free parameter. This way, we can select any $a$ between $-1$ and $1$, and the remaining parameters are found by solving the linear equation $b_0 z_0 + b_1 = z_0(z_0 - a)$, where $z\in\mathbb{C}$ is a chosen complex frequency point at which we want this equation to hold. Since there are two unknowns, we can write a second equation in which we want the derivative of each side of the equation also to be equal, yielding $b_0 = 2z_0 -a$. By replacing $b_0$ in the first equation, we obtain
\begin{equation*}
    \begin{split}
        z_0(2z_0 - a) + b_1 &= z_0(z_0 - a) \\
        b_1 &= -z_0^2.
    \end{split}
\end{equation*}
In the z-domain, the zero frequency corresponds to $z_0=1$. We choose to solve this equation for this point, hence we can replace $a$, $b_0$ and $b_1$  in Eq.~(\ref{eq:zequation}) which yields
\begin{equation}
    \Phi(z) = \frac{(2 - a)z - 1}{z - a}.
    \label{eq:aequation}
\end{equation}
This transfer function can then be converted to an associated state-space representation and used for time domain simulations with a sampling time $\Delta t$. The output of this system will then be $y_t \approx u_{t+\Delta t}$ for any input signal $u_t$. Simulating delays greater than the system sampling time is simply achieved by concatenating multiple times this predictor system. Here the delay has to be a multiple of the sampling time. This predictor can then be appended to the output of the digital controller $\mathcal{K}$.

\begin{figure}
    \centering
    \includegraphics{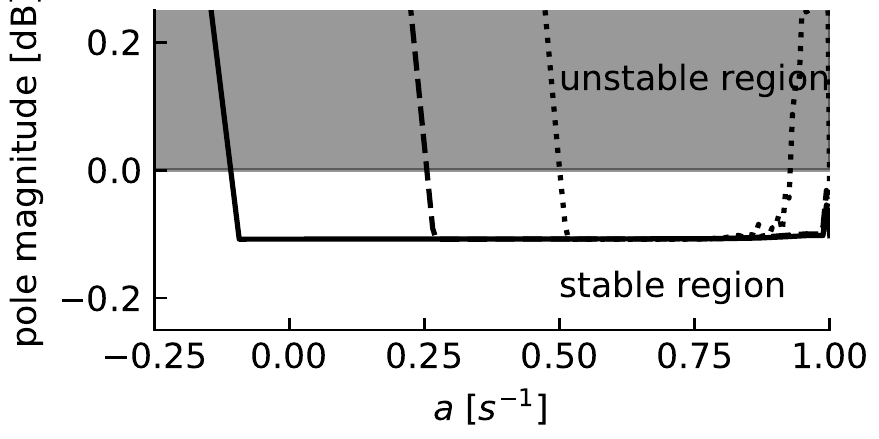}
    \caption{
        {\bf The predictor pole location affects the closed-loop stability.}
        The magnitude of the pole with the highest magnitude in the closed-loop transfer function parameterizes the stability of the closed-loop. Indeed, if this value is less than 0 dB, then all the poles of the closed-loop transfer function have a magnitude less than 0 dB, meaning that the system is stable. The system is unstable otherwise.
        Here the full curve, the dashed curve and the dotted curve correspond to predictors for delays of 3 ms, 5 ms and 10 ms, respectively.
        The higher the delay is, the lower is the size of the region of closed-loop stability for $a$.
    }
    \label{fig:predictor_stability}
\end{figure}

To avoid closed-loop instability, we must limit the amplitude of the feedback signal computed from the controller input signal. This amplitude is determined by the three systems $\mathcal{G}$, $\mathcal{H}$ and $\mathcal{K}$. Since $\mathcal{G}$ is defined by the system under study and  $\mathcal{H}$ is the chosen filter defining the desired modifications in the frequency distribution of the observed signal,  $\mathcal{\phi}$ (or equivalently parameter $a$) is the only degree of freedom. Figure~\ref{fig:predictor_stability} shows the region of closed-loop stability as a function of the predictor pole $a$ and the delay.

Because the predictor has a gain that is still slightly greater than one in the frequency ranges of interest, we reduce the weights of the filter $\mathcal{H}$ to compensate for the excess gain at the \textalpha~and \textgamma-peaks. To do this, we simply divide the weight of each band by the magnitude of the predictor system evaluated at the band's natural frequency. This reduces the errors in the closed-loop transfer function in the \textalpha~and \textgamma-ranges.

\begin{figure}[ht!]
\begin{center}
\includegraphics[width=\linewidth]{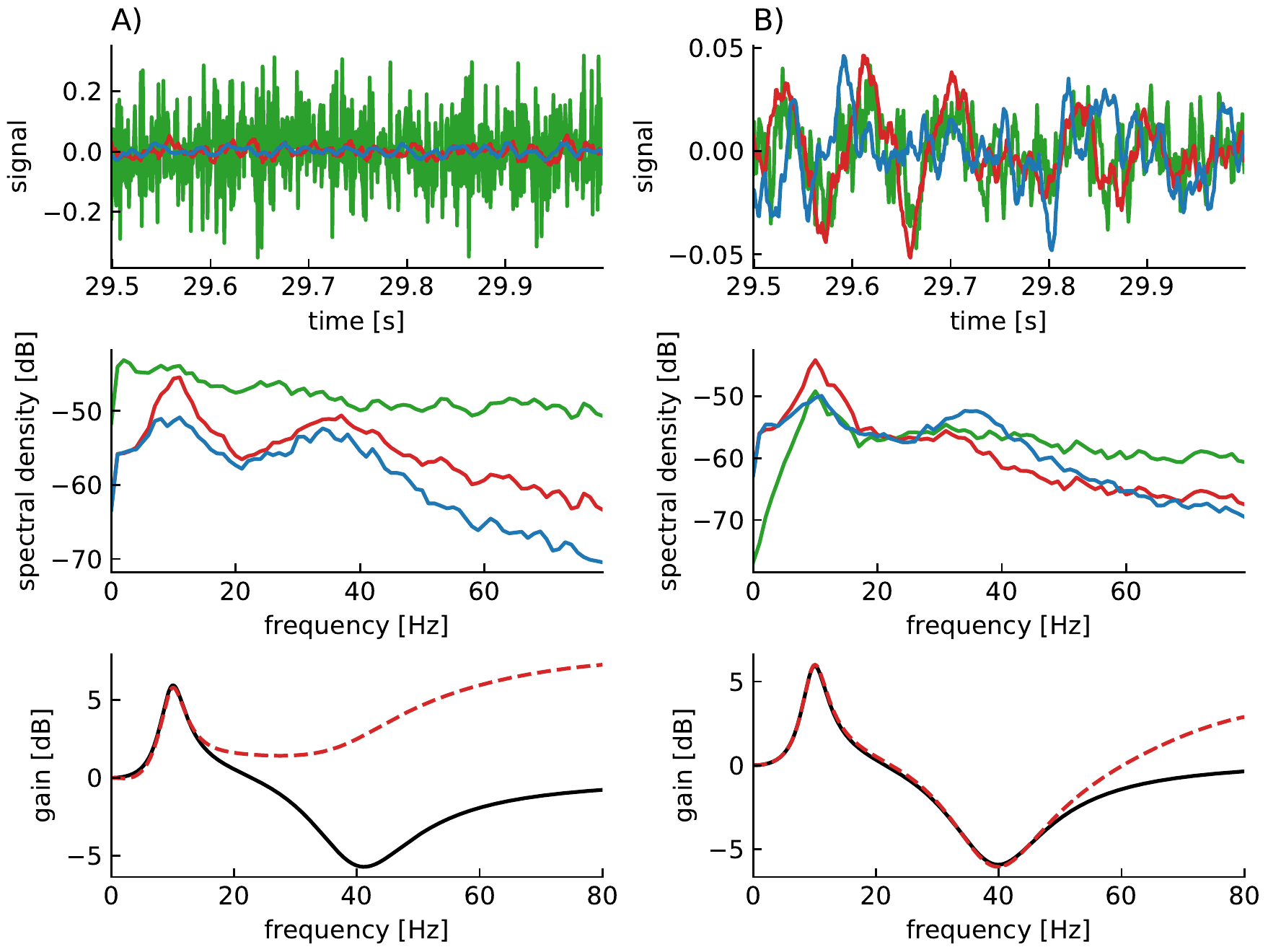}
\end{center}
\caption{
    {\bf Model-based closed-loop neurostimulation with delay compensation successfully decreases gamma activity while reference signal-based control with Smith predictor fails.}
    {\bf A)} Simulation data of the reference signal-based control design with Smith predictor.
    {\bf B)} Simulation data of the model-based control design with delay compensation.
    The upper panels show the time series of the resting state activity signal $y_0$ (blue) and the closed-loop output signal $y$ (red) and the input current $u$ (green). The amplitude of the stimulation current is much larger for reference signal-based control than for model-based control. The center panels show spectral densities of the resting state activity signal $y_0$ (blue), the closed-loop output signal $y$ (red) and the input current $u$ (green). The activity is increased in the alpha range and decreased in the gamma range for model-based control, however, is increased everywhere for reference signal-based control. The spectral density of the input current is again much larger for reference signal-based control than for model-based control.  The lower panels show the spectral density gain from $y_0$ to $y$ of the closed-loop systems. The dashed red curve is computed from the closed-loop transfer function and the black curve is the target curve computed from the transfer function $1 + H(s)$. We see that the implemented closed-loop applies the correct modifications in alpha and gamma ranges for model-based control but not for reference signal-based control where the error is large for frequencies above the alpha range. The conduction delay is $5$ms and the value of the parameter $a$ in the delay compensation scheme is chosen to $a=0.55$
}
\label{fig:compare}
\end{figure}

Figure~\ref{fig:compare}(B) shows results combining the model estimation by vector fitting and the delay compensation. The proposed closed-loop control yields an increase in \textalpha-power and a decrease in \textgamma-power according to the employed target filter $\mathcal{H}$. The application of a conventional reference signal control and Smith predictor for delay compensation (Fig.~\ref{fig:compare}(A)) does not yield a reduction of higher \textgamma-frequency activity. This can also be seen in Fig.~\ref{fig:compare}(bottom panel), showing that the proposed scheme adapts much better to the target gain function than the reference signal control scheme. Generally, both methods fail to adapt well to very high-frequencies (details not shown).

\subsubsection{Accuracy}
\begin{figure}[ht!]
\begin{center}
    \includegraphics[width=.5\linewidth]{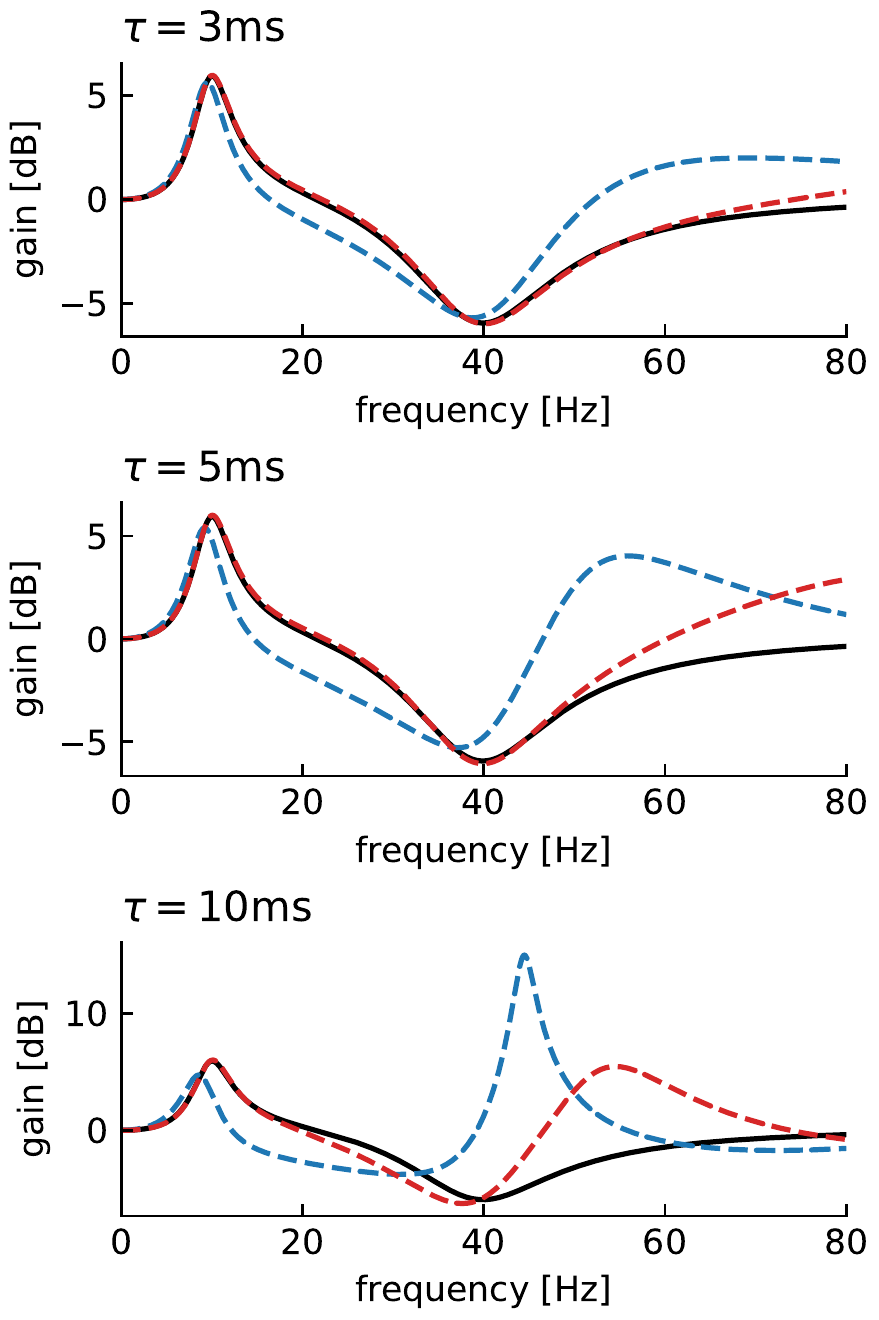}
\end{center}
\caption{
    {\bf Delay decreases the accuracy of the closed-loop transfer function}
    For uncompensated delay (dashed blue curve), the closed-loop transfer function significantly deviates from the target transfer function defined as $1 + H(s)$ (black curve). Delay compensation (dashed red curve) reduces the deviation from the target transfer function in the \textalpha- and \textgamma-frequency range for delays of $3$ms and $5$ms. However, the error is still large in the \textgamma-range for a delay of $10$ms.
}
\label{fig:delay}
\end{figure}

\subsubsection{Stability}
As discussed earlier, delay compensation can destabilize the closed-loop system depending on the parameters of its components. However, if the correct predictor pole is chosen based on Fig.~\ref{fig:predictor_stability}, the closed-loop will remain stable. These values are computed under the assumption that there are no model estimation errors. If we take into account the inaccuracies in the fitted brain model compared to the original brain model, extra gain can add up in the feedback signal, introducing again the risk of destabilizing the closed-loop. This is trickier to solve, as we assume here that in a real experimental setup that, it is very difficult to reduce these remaining errors further by the method proposed. Hence the solution is either to simply reduce the amplitude of the spectral density modification that we want to apply by reducing the amplitude of the transfer function of filter $\mathcal{H}$, or to reduce the amplitude of the predictor $\mathcal{\Phi}$ reducing its accuracy and possibly increasing delay errors. In any case, the inaccuracies in the estimated brain model create errors in the closed-loop transfer function regardless of the delay.

\subsection{Application to cortico-thalamic circuit model}
\begin{figure}[ht!]
\begin{center}
    \includegraphics[width=\linewidth]{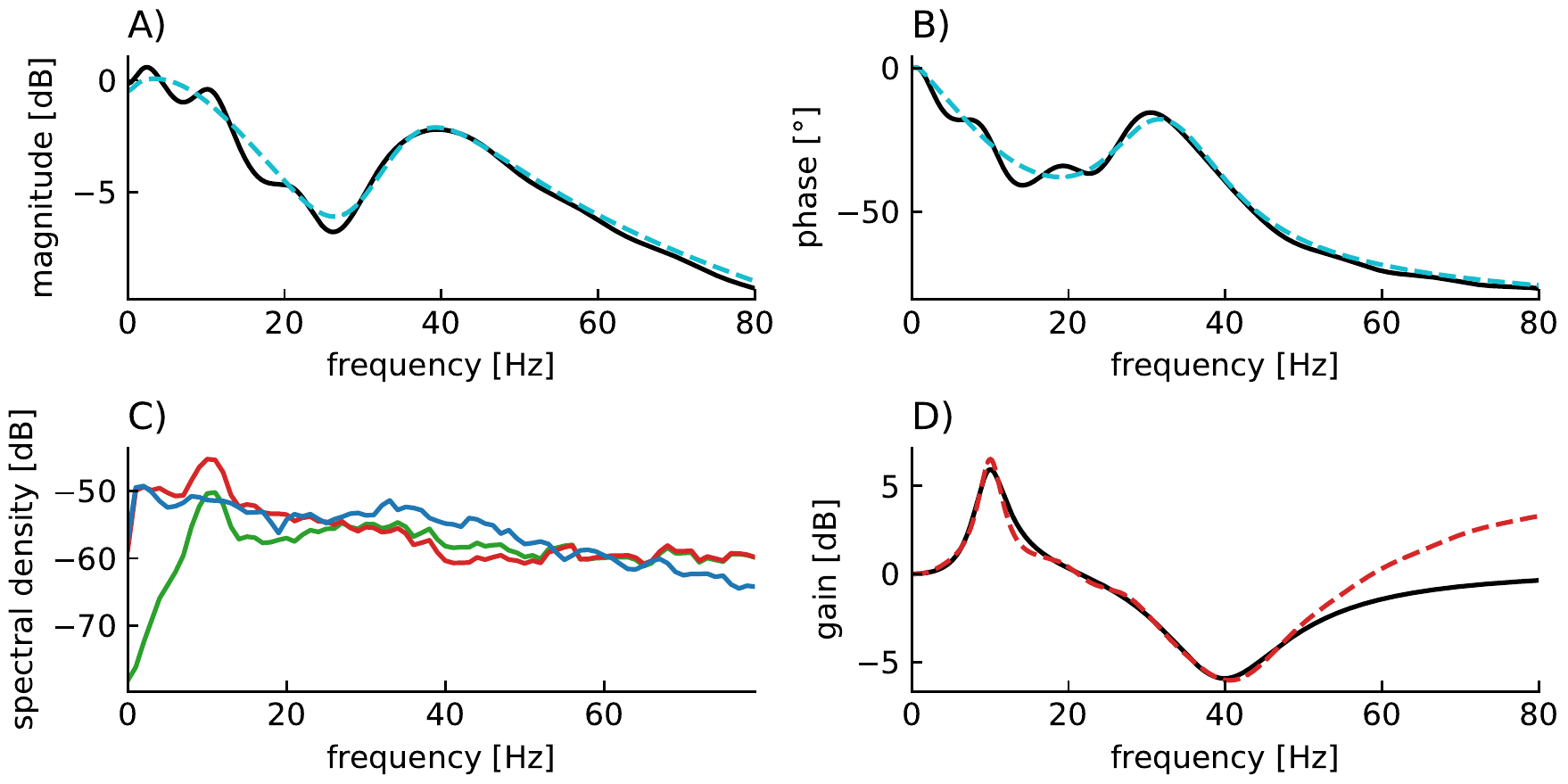}
\end{center}
\caption{
    {\bf Fitted model-based control using the cortico-thalamic brain model successfully reproduces the target transfer function in the frequency domains of interest.}
    {\bf A)} Magnitude of the fitted brain model transfer function (dashed cyan) compared to the magnitude of the original cortico-thalamic brain model transfer function (black).
    {\bf B)} Phase shift of the fitted transfer function (dashed cyan) compared to the magnitude of the original transfer function (black). {\bf C)} Spectral densities of the rest state activity signal (blue), the stimulated brain output (red) and the stimulation signal (green). {\bf D)} Closed-loop transfer function (dashed red), compared to the target transfer function $1 + H(s)$ (black).
} 
\label{fig:ct-closed-loop}
\end{figure}

To extend the analysis to a biologically more realistic model, we employed a nonlinear cortico-thalamic brain model (cf. section~\ref{subsebsection_cct}). Fitting a linear transfer function to the brain model activity as described above, we found a good accordance of fitted and original model as can be seen in Fig.~\ref{fig:ct-closed-loop}A),B). Small deviations in the gain and the phase resulted from the internal delay in the brain model and its non-linearity. Indeed, the magnitude vector fitting algorithm does not reproduce this delay but instead synthesizes a linear system that has no delay but still approximates well the transfer function of the original model. Nonetheless, the non-linearity of this model can also decrease the accuracy of the fitting, as we are trying to represent a non-linear input response model by a linear one. However, this effect is only seen when the current is large enough for the non-linear part of the response to be significant.

In fact, the model-based control enhances \textalpha-activity and diminishes \textgamma-activity in good accordance to the imposed filter $\mathcal{H}$ (see Fig.~\ref{fig:ct-closed-loop}C)). This can also be seen in the closed-loop transfer function, which corresponds well to the target transfer function (see Fig.~\ref{fig:ct-closed-loop}D)) for small and medium frequencies. The closed-loop transfer function deviates from the target transfer function for large frequencies beyond the \textgamma-frequency range. This results from the employed conduction delay.

\begin{figure}[ht!]
\begin{center}
    \includegraphics[width=\linewidth]{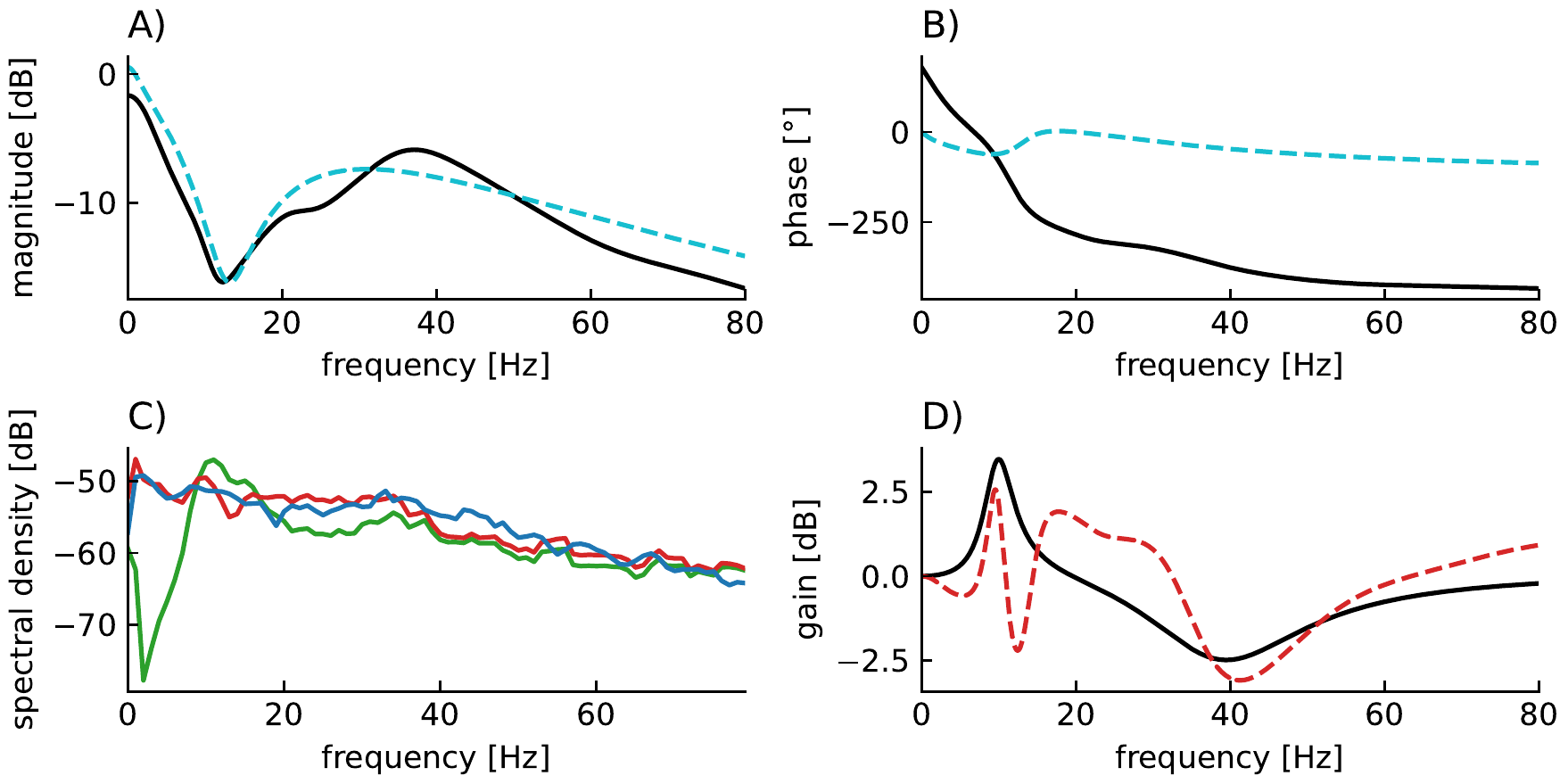}
\end{center}
\caption{
    {\bf Reticulum stimulation yields incorrect closed-loop gain in low-frequency ranges.}
    {\bf A)} Magnitude of the fitted brain model transfer function (dashed cyan) compared to the magnitude of the original cortico-thalamic brain model transfer function (black).
    {\bf B)} Phase shift of the fitted transfer function (dashed cyan) compared to the magnitude of the original transfer function (black). {\bf C)} Spectral densities of the rest state activity signal (blue), the stimulated brain output (red) and the stimulation signal (green). {\bf D)} Closed-loop transfer function (dashed red), compared to the target transfer function $1 + H(s)$ (black).
}
\label{fig:ctretstim}
\end{figure}

To elucidate better the functions of the different elements of the proposed method, we applied a second closed-loop setup, where the neurostimulation input was applied to the first three layers of the cortex modeled by $u$ and $v$ and to the reticulum modeled by $V_{ret}$ (Fig.~\ref{fig:ctretstim}). In this setting, the response in the high-frequency ranges are mainly produced by the cortex, while the response in low-frequency ranges originates mainly from the reticulum and the thalamic relay structure, with a gap approximately between $10$Hz and $20$Hz. The weak response between 10 Hz and 20 Hz observable cf. Fig.~\ref{fig:ctretstim}A is compensated by the controller, which produces a high magnitude stimulation in the closed-loop for these frequencies cf. Fig.~\ref{fig:ctretstim}C. The second consequence is the inaccuracy of the closed-loop output in the low-frequency ranges, this is caused by the rather long cortico-thalamic internal delay. This delay yields a larger phase shift at low-frequencies and originates from the fact that we observe signals in the cortex, but stimulate in the reticulum.

\section{Discussion}
\label{sec:discussion}

The goal of the proposed method was to design a delayed closed-loop control method to apply defined modifications to the spectral distribution of an observed signal, such as EEG or LFP. The presented work explicitly describes all the steps needed to build a delayed closed-loop neurostimulation setup to restore the physiological brain state of a patient \cite{hebb2014}. Since the controller is modeled as a linear time-invariant system, its implementation is lightweight, straightforward, and easily applicable in most embedded systems. Applications to a simple neural populations model (Fig.~\ref{fig:compare}) and to a biologically plausible cortico-thalamic feedback system (Fig.~\ref{fig:ct-closed-loop} and \ref{fig:ctretstim}) demonstrate its elements and their impact on the control performance.

\subsection*{Main contributions}

\subsubsection*{Model estimation}
We assume resting state activity signal driven by noise, when no neurostimulation is applied. Injecting a stimulation creates an additional response that adds to the resting state. Consequently, both the resting state signal and response signal can be observed separately in experimental practice and they serve to estimate a linear state-space model as outlined in section~\ref{subsec_modelestimation}. This approach is successful for both simplified linear models (cf. Figs.~\ref{fig:psdfit},\ref{fig:noise_level}) and neurophsysiological realistic nonlinear models (cf. Fig.~\ref{fig:ct-closed-loop}). This approximation is suitable for nonlinear systems whose dynamics evolve close to a stationary state. Several studies have already exposed evidence confirming that the measured brain dynamics behave mostly linearly at macroscopic scales \cite{Liu2010}, \cite{Popivanov1996}. Moreover, in the case of the brain response to small neurostimulation input, our assumption of the linear brain response is supported by results of \cite{Kim2016}. The authors of this study measured the controllability Gramian of their brain model with nonlinear sigmoid transfer function, similar to the cortico-thalamic brain model \cite{riedinger2022} used in this paper. If the system exhibits nonlinear dynamics far from any linear approximation, such as bistable dynamics and chaotic evolution, the proposed vector fitting technique may yield a too large model error and thus instability of the closed-loop feedback. The hypothesis of macroscopically linear dynamics has also recently been tested against various nonlinear models \cite{Nozari2020}. While that work  included fitting methods for both linear and nonlinear brain models, our work chose the paradigm of purely frequency domain model fitting with the magnitude vector fitting algorithm \cite{de2010robust} and applied it to the brain input response system, which we could isolate thanks to a simple open-loop neurostimulation setup. While models have already been studied in application to neurostimulation \cite{Modolo2011}, \cite{Wagner2014}, we propose a straightforward black box modeling approach that is directly usable for adaptive closed-loop neurostimulation, and is technically applicable easily for each individual patients before any closed-loop neurostimulation sessions.

\subsubsection*{Delay compensation}
Conduction delays of a few milliseconds in the transmission between observation and stimulation may be negligible in systems evolving on time scales of seconds or longer, but may play an important role in neural systems. Our study demonstrates that such feedback delays may introduce control errors and we show how these errors can be avoided by a novel delay compensation method (section~\ref{sec:results:delay_compensation}). Application to the linear model~\eqref{fig:compare} demonstrated its superior performance compared to a conventional delay compensation method. Delay compensating systems have already been described in other work \cite{Hosseini2019}, \cite{Guo2004}. However, we used a design primarily focused on the correction of a gain error in the closed-loop transfer function, whereas the majority of the current research is based on time domain criterion and stability enforcement \cite{Ledva2017}, \cite{Sonmez2015}. The methods performance, i.e. how well the total gain function fits to the pre-defined transfer function, is good for low-frequencies but weakens for frequencies exceeding a limit frequency. Note that frequency domain compensation has also already been achieved, notably via delay equalizers \cite{Podilchak2009}. However, this would restrict the frequency range in which the delay is compensated, and create additional errors in the surrounding frequencies. Other designs include filters with negative group delays, however their applications are limited to band limited input signals \cite{Bukhman2004}, \cite{Voss2017}. The predictor design we presented also relies on negative group delay, enabling delay compensation in a large frequency band, while still being applicable to the brain EEG, which is inherently not band limited, because of the noise. Nonetheless, while our predictor design allows to significantly decreases the delay errors in the closed-loop transfer function, the delay still imposes a limit on the controllable frequency range. The larger the delay, the smaller is this limit frequency. Low performance may induce instability in the feedback loop~\cite{Mirkin2005} and thus should be avoided. A corresponding stability criteria has been proposed, cf. Fig.~\ref{fig:predictor_stability}. Better predictor designs could allow better performance of the closed-loop system for larger delays. The improvement of the accuracy of our closed-loop neurostimulation setup by building more efficient predictor designs is in progress and we refer the reader to future work.

\subsection*{Limits of our methodology}

\subsubsection*{Experimental stimulation parameters and safety}
Experimental stimulation protocols have to ensure the subjects safety~\cite{Ko} and thus avoide stimulus-induced health risks and complications. For instance, tDCS may be administered for a duration of 60 minutes and a maximum current of 4 mA without yielding health risks. However, parameters beyond these limits may yield adverse effects in subjects, such as skin lesions similar to burns and mania or hypomania in patients with depression~\cite{MATSUMOTO201719}. The proposed method does not limit the stimulation duration {\em per se}, but of course the duration can be chosen accordingly without constrating the method. The method adapts the systems brain rhythms to the target rhythms very rapidly on a time scale of less than a second and hence permits rather short stimulation duration longer than a second.

Moreover, the proposed method does not specify absolute stimulation current magnitude applied. The impact of stimulation at certain magnitudes depends heavily on the stimulation type. In tDCS, anodal stimulation with positive currents have a different impact as cathodal stimulation with negative currents. In addition, currents are thought to have to pass a certain threshold to yield a measurable effect. In tACS~\cite{Moliadze}, stimulating in the \textalpha-frequency range large and small magnitudes yield excitation and inhibition, respectively, while intermediate magnitudes yield weak effects. Stimulating with a range of frequencies, as in tRNS~\cite{Potok}, a 1mA peak-to-peak amplitude for 10 minutes stimulation duration does not yield adverse effects. We conclude that it is not straight-forward to decide which stimulation magnitude applied in the presented method would be  safe for human subjects, since the stimulation signal is neither constant, single frequency oscillation nor random noise. In sum, we argue that a maximum peak-to-peak amplitude of 1mA for few tens of minutes may not yield adverse effects, but still may evoke a measurable impact on observations and the brain state. Of course, future experimental studies will gain deeper insights.

\subsubsection*{Model internal delay}

The internal delay in the brain is not reproducible by the magnitude vector fitting algorithm, which relies on the time invariance of the signals. Hence, this will cause errors in the transfer function of the fitted model (cf. Fig.~\ref{fig:ct-closed-loop}) that are larger for higher contribution of the delay in the output, cf. Fig.~\ref{fig:ctretstim}. To limit this effect, we must minimize the delay between the application of the neurostimulation input and the measurement of the response to this input as much as possible by taking into account the delay between the different brain regions.

\subsubsection*{Estimating the closed-loop delay}

For delay compensation, in this paper, we assumed that we know the conduction delay in the closed-loop. However, although it is a single constant parameter, we would need a method to measure it for a real closed-loop neurostimulation setup. A straightforward way to do this would be to inject any current into the plant and measure the time lag between the moment at which we inject the input current and the moment at which we measure the output signal. This estimated delay would then correspond to the total closed-loop delay except for the computation delay of the digital controller $\mathcal{K}$. This computation delay can be easily measured with the same software used for computation, as it corresponds to the delay needed to perform constant-size matrix multiplications. Moreover, several methods have already been developed to estimate the conduction delays in linear systems \cite{Schier1997}, \cite{Dudarenko2014}.

\subsubsection*{Direct input current measurements}

One of the main challenges to solve for closed-loop neurostimulation is the elimination of direct transmission artifacts from the measured EEG signal \cite{iturrate2018}. Indeed, when measuring the plant output signal, a portion of the measured signal might be a direct measurement of the input current without any influence from the brain dynamics. In the ideal case, one intends to minimize the contribution of the stimulation input to the observed signal since it would mean that the measured EEG signal does not fully correspond to the brain activity. Hence, reading the EEG of the patient would be more difficult for the user of our closed-loop setup, and the contribution of the brain dynamics to the closed-loop would be smaller. A simple solution to this problem is discussed further below.

\subsection*{Perspectives}
The control proposed allows to perform accurate frequency shaping of the systems' activity spectral distribution. However, this approach is limited to linear models of the brain stimulation response. This may be disadvantageous if the systems dynamics exhibit nonlinear behavior (see e.g.~\cite{Hutt_meta}) as we want to  represent the brain dynamics realistically. Furthermore, in real-case scenarios, we would also have to take into account the noise in the acquisition of the signal by the sensor and in the application of the input signal by the actuator.

\subsubsection*{Filtering out direct input current measurements}
\label{sec:remove_interaction}

Filtering out the direct input current measurements is achievable with our setup removing the strictly proper system requirement while using the magnitude vector fitting algorithm to measure the brain input response. In other words, while fitting the brain input response system, we want the fitted model to be able to contain a direct transmission term corresponding to the direct current measurement. Hence, if the real plant input response contains a significant direct transmission term, it will be identified by the magnitude vector fitting algorithm when synthesizing the estimated plant input response. The second step is them simply to substract the feedtrough term multiplied by the input current to the plant output signal. Thus, the remaining part of the signal would only correspond to the brain dynamics.

\subsubsection*{Application to multiple inputs multiple outputs plants}

For now, we only focused on plant with a signal input signal and a single output signal. However, in a real setup, the EEG measurement is typically composed of multiple channels corresponding to different electrodes. This can also be true for the neurostimulation device. For example, with electric current stimulation, we can inject multiple signals  using multiple electrodes. This can be simply solved by feeding a single input to each input channel and summing each output to a single output channel. However, when we separate the different channels, we can have more control over each individuals output channels. When we have multiple inputs and output, the plant is then a Multiple-Inputs Multiple-Outputs (MIMO) system. Everything developed in this paper is generalizable to MIMO systems, with one caveat: when solving Eq.~(\ref{eqn:K}), a unique solution only exists if the system has as more outputs than it has inputs. The user can always ensure this, by using as many neurostimulation input channels than there are EEG output channels. In this generalized setup, we can also define the filter $\mathcal{H}$ to apply different modifications to each output channel.

\subsubsection*{Neurostimulation effects on larger time scales}

Our method relies only on the short term dynamics of the brain, using signal feedback and delay compensation to produce an adaptive stimulation current and obtain the desired EEG frequency distribution. However, more traditional neurostimulation techniques rely on the long term dynamics of neural plasticity, which is not modeled in the brain models we use in this paper. Long term brain adaptation to neurostimulation could cause the EEG frequency distribution to diverge from the desired frequency distribution after several minutes of stimulation. This effect could be compensated either by reiterating the model identification step and performing neurostimulation again, or  by adjusting the weight of the filter $\mathcal{H}$ according to the observed changes in real-time. Incorporating the effect of neural plasticity in the brain models would allow our method to produce predictable and durable modification to the EEG frequency distribution, even after we stop the stimulation.

\section*{Conflict of Interest Statement}

The authors declare that the research was conducted in the absence of any commercial or financial relationships that could be construed as a potential conflict of interest.

\section*{Author Contributions}

TW, AH and MD contributed to the development of the methods presented in this study. TW produced the source code used for the simulations. JR and AH wrote the introduction section. TW and AH wrote the other sections of the manuscript. All the authors read and approved the submitted version.

\section*{Funding}
This research was funded by Inria in the "Action Exploratoire" project {\em A/D Drugs}.

\section*{Data Availability Statement}
The source code used for the simulation results can be freely accessed here: \url{https://github.com/Thomas-Wahl/neuroclodec}

\bibliographystyle{unsrt}
\bibliography{main}
            
\clearpage

\appendix

\section*{Cortico-thalamic brain model details}

The differential equation system (\ref{eq:ct}) develops as

\begin{equation}
    \resizebox{\textwidth}{!}{$
        \begin{aligned}
            \tau_e\frac{dV_e(t)}{dt} &= -V_e(t) + F_e T_c(V_e(t) - V_i(t)) + F_{ct}T_{th}(V_{th,e}(t) - V_{th,i}(t)) + F_{ccx}S_e(v(t)) + \mu_e + I_e + \xi_e(t) + b_1 u(t) \\
            \tau_i\frac{dV_i(t)}{dt} &= -V_i(t) + F_i T_c(V_e(t) - V_i(t)) + \mu_i + I_i + \xi_i(t) + b_2 u(t) \\
            \tau_{th,e}\frac{dV_{th,e}(t)}{dt} &= -V_{th,e}(t) + F_{tc}T_c(V_e(t - \tau) - V_i(t - \tau)) + \mu_{th,e} + \xi_{th,e}(t) \\
            \tau_{th,i}\frac{dV_{th,i}(t)}{dt} &= -V_{th,i}(t) + F_{tr}T_{ret}(V_{ret}(t)) + \mu_{th,i} + \xi_{th,i}(t) \\
            \tau_{ret}\frac{dV_{ret}(t)}{dt} &= -V_{ret}(t) + F_{rt}T_{th}(V_{th,e}(t) - V_{th,i}(t)) + F_{rc}T_c(V_e(t - \tau) - V_i(t - \tau)) + \mu_{ret} + \xi_{ret}(t) \\
            \tau_{ce}\frac{dv(t)}{dt} &= -v(t) + F_{cx}S_e(v(t)) - M_{cx}S_i(w(t)) + M_{cx,th}T_{th}(V_{th,e}(t - \tau) - V_{th,i}(t - \tau)) + \mu_{ce} + I_{ce} + \xi_{ce}(t) + b_3 u(t) \\
            \tau_{ci}\frac{dw(t)}{dt} &= -w(t) - F_{cx}S_i(w(t)) + M_{cx}S_e(v(t)) + \mu_{ci} + I_{ci} + \xi_{ci}(t) + b_4 u(t)
        \end{aligned}$
    }
    \label{eq:corticothalamic}
\end{equation}
where the transfer functions are defined as
\begin{equation}
    \begin{split}
        T_m(x) &= \frac{1}{2}\left(1 - \mathrm{erf}\left(-\frac{x}{\sqrt{2}\sigma_m}\right)\right) \\
        S_m(x) &= \frac{1}{2}\left(1 - \mathrm{erf}\left(-\frac{x}{\sqrt{2}\sigma_{cm}}\right)\right).
    \end{split}
    \label{eq:cttf}
\end{equation}
The $\xi_x$ terms represent the driving noises, which are uncorrelated Gaussian noise defined as
\begin{equation*}
    \langle\xi_x(t)\rangle = 0\quad,\quad \langle\xi_x(t)\xi_y(t')\rangle = \frac{Q_x}{N}\delta_{xy}\delta(t-t'),
\end{equation*}
with $x = e, i, (th,e), (th,i), ret, ce, ci$. The variances in Eq.~(\ref{eq:cttf}) are defined as
\begin{eqnarray*}
    \sigma_c^2 = \frac{Q_e}{\tau_e} + \frac{Q_i}{\tau_i} ,&\quad \sigma_{th}^2 = \frac{Q_{th,e}}{\tau_{th,e}} + \frac{Q_{th,i}}{\tau_{th,i}} ,&\quad \sigma_{ret}^2 = \frac{Q_{ret}}{\tau_{ret}} \\
    \sigma_{ce}^2 = \frac{Q_{ce}}{\tau_{ce}} ,&\quad \sigma_{ci}^2 = \frac{Q_{ci}}{\tau_{ci}}
\end{eqnarray*}

All the parameters are given in Table \ref{table:ctparameters}.

\begin{table}
    \centering
    \begin{tabular}{lll}
        \hline\noalign{\smallskip}
        parameter & description & value  \\
        \noalign{\smallskip}\hline\noalign{\smallskip}
        $\tau_e$      & exc. decay time (infragranular)    & 10 ms \\
        $\tau_i$      & inh. decay time (infragranular)    & 50 ms \\
        $\tau_{th,e}$ & exc. decay time (relay)            &  5 ms \\
        $\tau_{th,i}$ & inh. decay time (relay)            & 30 ms \\
        $\tau_{ret}$  & exc. decay time (reticular)        &  8 ms \\
        $\tau_{ce}$   & exc. decay time (supragranular)    &  5 ms \\
        $\tau_{ci}$   & inh. decay time (supragranular)    & 20 ms \\
        $\tau$        & cortico-thalamic propagation delay & 40 ms \\
        $F_e$         & exc. synaptic strength                                         & 1.0  \\
        $F_i$         & inh. synaptic strength                                         & 2.0  \\
        $F_{ct}$      & synaptic strength (relay $\rightarrow$ cortex)                 & 1.2  \\
        $F_{tc}$      & synaptic strength (cortex $\rightarrow$ relay))                & 1.0  \\
        $F_{tr}$      & synaptic strength (reticular $\rightarrow$ relay)              & 1.0  \\
        $F_{rt}$      & synaptic strength (relay $\rightarrow$ reticular)              & 0.3  \\
        $F_{rc}$      & synaptic strength (cortex $\rightarrow$ reticular)             & 0.6  \\
        $F_{cx}$         & synaptic strength (exc. $\rightarrow$ exc.)                 & 2.18 \\
        $M_{cx}$         & synaptic strength (inh. $\rightarrow$ exc.)                 & 3.88 \\
        $F_{ccx}$     & synaptic strength (supragranular $\rightarrow$ infragranular)  & 0.05 \\
        $F_{cx,th}$   & synaptic strength (thalamic relay $\rightarrow$ supragranular) & 0.1  \\
        $\mu_e$       & exc. noise input (infragranular)   & 0.1  \\
        $\mu_i$       & inh. noise input (infragranular)   & 0.0  \\
        $\mu_{th,e}$  & exc. noise input (relay)           & 1.3  \\
        $\mu_{th,i}$  & inh. noise input (realy)           & 1.0  \\
        $\mu_{ret}$   & exc. noise input (reticular)       & 0.0  \\
        $\mu_{ce}$    & exc. noise input (supragranular)   & 0.05 \\
        $\mu_{ci}$    & inh. noise input (supragranular)   & 0.05 \\
        $I_e$         & exc. resting input (infragranular) & 2.7  \\
        $I_i$         & inh. resting input (infragranular) & 1.7  \\
        $I_{ce}$      & exc. resting input (supragranular) & 1.1  \\
        $I_{ci}$      & inh. resting input (supragranular) & 0.4  \\
        $Q_e$      & exc. input noise variance (infragranular) & $3\times10^{-5}$    \\
        $Q_i$      & inh. input noise variance (infragranular) & $0.001$             \\
        $Q_{th,e}$ & exc. input noise variance (relay)         & $2.5\times10^{-6}$  \\
        $Q_{th,i}$ & inh. input noise variance (relay)         & $12.6\times10^{-6}$ \\
        $Q_{ret}$  & exc. input noise variance (reticular)     & $10.9\times10^{-6}$ \\
        $Q_{ce}$   & exc. input noise (supragranular)          & $2\times10^{-5}$    \\
        $Q_{ci}$   & inh. input noise (supragranular)          & $8\times10^{-5}$    \\
        $N$ & number of neurons & 1000 \\
        $b_{1, 2, 3, 4}$ & input coupling constants & 1 \\
        $c_1$ & observation coefficient (supragranular) & 0.3 \\
        $c_3$ & observation coefficient (infragranular) & 1   \\
        \noalign{\smallskip}\hline
    \end{tabular}
    \caption{
        {\bf Parameter set of model (\ref{eq:corticothalamic}).}
        The choice of parameters is for the most part based on the paper in which it was developed~\cite{riedinger2022}.
    }
    \label{table:ctparameters}
\end{table}

\end{document}